\documentclass[3p,times,twocolumn]{elsarticle}
 \biboptions{comma,sort&compress}
\usepackage{graphicx}
\usepackage{here}
\usepackage{ecrc}
\usepackage{amsmath,amsbsy,mathtools}
\usepackage{latexsym}
\usepackage{amssymb}
\usepackage{bbm}
\def\vec#1{\boldsymbol{#1}}
\def\overl#1{\mskip2mu\overline{\mskip -2mu#1}{}}

\mathchardef\mhyphen="2D

\volume{00}

\firstpage{1}

\journalname{Nuclear and Particle Physics Proceedings}

\runauth{}

\jid{nppp}

\jnltitlelogo{Nuclear and Particle Physics Proceedings}

\usepackage{amssymb,amsmath,amsbsy}
\usepackage[figuresright]{rotating}

\begin{document}

\begin{frontmatter}

%%
%%%%%%%%%%%%%%%%%%%%%%%%%%%%%%%%%%%%%%%%%%%%%%%%%
\title{Fully-heavy tetraquarks and other heavy multiquarks
 $^*$}
 % \corref{cor0}}
 \cortext[cor0]{Talk given at 23rd International Conference in Quantum Chromodynamics (QCD 20,  35 years later),  29 June - 3 July 2015, Montpellier, France}
 \author[label1]{Jean-Marc Richard}
%  \cortext[cor0]{FAPESP CNPq-Brasil PhD student fellow.}
\ead{j-m.richard@ipnl.in2p3.fr}
\address[label1]{Institut de Physique des 2 Infinis de Lyon, Université de Lyon, UCBL-CNRS-IN2P3, 4, rue Enrico Fermi, Villeurbanne, France}
\pagestyle{myheadings}
\markright{ }
\begin{abstract}
A review is presented of some discussions about the fully-heavy tetraquarks made of two charmed quarks and two anticharmed antiquarks,  and other multiquark states involving some heavy flavors.
\end{abstract}
% \begin{document}
\begin{keyword} 
heavy quarks \sep multiquarks \sep tetraquarks \sep pentaquarks \sep hexaquarks
\end{keyword}
\end{frontmatter}
%%%%%%%%%%%%
%\vspace*{-1.5cm}
\section{Introduction}\label{se:intro}
Exotic hadrons is a delicate subject, and it is even difficult to admit that one is interested in the subject: \textsl{Do not tell my mother that I am working on exotic hadrons, she thinks I am a pianist in a brothel}\footnote{Adapted from \cite{seguela1979ne}}
%the book entitled ``Ne dites pas à ma m\`ere que je suis dans la publicit\'e, elle me croit pianiste dans un bordel'', by Jacques S\'egu\'ela.}. 
There have been, indeed, embarrassing ups and downs, both on the experiment and theory sides, with an interlaced share of responsibilities. See, e.g., \cite{Richard:2016eis,Ali:2019roi,Brambilla:2019esw,Richard:2020uan}, for a sketch of the history.  

For instance, erroneous amplitude analyzes have led to speculate on the existence of exotic hyperons with positive strangeness, the so-called $Z$-baryons. It is nowadays acknowledged  that a serious understanding of the $KN$ system requires a minimal amount of spin-dependent measurements. Not surprisingly, the same authors have carried out somewhat questionable analyzes of the nucleon-antinucleon annihilation into two pseudoscalar mesons, tentatively leading to the existence of broad baryonium resonances.

More recently, there has been debates on the predictions of simple models. One can take, for instance, a simple pairwise, chromoelectric model
\begin{equation}\label{eq:HN}
 H=\sum_{i=1}^N\frac{\vec p_i^2}{2\,M}- \frac{3}{16} \sum_{i<j}\tilde\lambda_i.\tilde\lambda_j\,v(r_{ij})~,
\end{equation}
where $M$ is the mass of the quark in the $QQ\overl Q\overl Q$ system, $\tilde\lambda$, the color operator, and the factor such that the interaction is mediated by the exchange of a color octet, $v(r)$ being the quarkonium potential. The question is whether or not the ground state lies below the threshold for spontaneous dissociation into two $Q\overl Q$ singlets, and there are conflicting answers in the literature. If one cannot address and resolve safely that question, who will believe that our community can handle more ambitious approaches?

We should thus advocate for a careful treatment of the multiquark systems, even in the simplistic schemes. It took decades to the community of atomic and nuclear physics to device powerful tools such as the Faddeev-Yakubosky equations, or the hyperspherical expansion, or a variational calculation based on the correlated Gaussian (for refs., see, e.g., \cite{Richard:2020uan}), thus enabling us to understand delicate structures such as the Efimov states, the positronium molecule, of the ${}_\Lambda^3\mathrm{H}$ hypernucleus. Who will reasonably believe that for an $N$-body, color-singlet, quark system governed by Eq.~\eqref{eq:HN}, one can  solve for 2-body subsystems, iteratively, and never enter the cumbersome techniques of 3-body or 4-body dynamics? In other words, is any baryon made of a diquark and a quark, and any tetraquark a 2-body system made of a diquark and an antidiquark? A related question deals with the terminology: in the 70s, Chan H.M., and others, carefully named ``diquonium'' the very peculiar four-quark configuration with a strong clustering of the two quarks and of the two antiquarks, leaving room for other shapes \cite{Chan:1978nk}. Nowadays, the diquark lobby has imposed a cheeky confusion between tetraquark and diquonium. 
\section{Chromo-harmonic confinement}
Let us consider a special case of \eqref{eq:HN} where $M=1$ and $v(r)=r^2$. Introducing the usual Jacobi variables
\begin{equation}
 \vec x=\vec r_2-\vec r_1~,\quad \vec y=(2\,\vec r_3-\vec r_1-\vec r_1)/\sqrt3~,
\end{equation}
and their conjugate momenta, the intrinsic part of $H$ reads
\begin{equation}\label{eq:HN3}
 H_\text{int}=\vec{p}_x^2+\frac34\,\vec x^2+\vec{p}_y^2+\frac34\,\vec y^2~,
\end{equation}
while in a naive diquark approach, one first solves for the diquark with a potential $\vec x^2/2$, and then the diquark-quark system with a potential $3\,\vec y^2/2$. So the $\vec x$ part of the potential is lowered by 50\%, and the corresponding energy by about 25\%. It is easily seen that the comparison also holds for any quark-mass configuration $MMm$, whatever the mass ratio $M/m$ is adopted. 

For tetraquarks, there are two independent color configuration to build a color singlet. We adopt here the notation of \cite{Chan:1978nk}, namely $T=\bar 3\mhyphen 3$ and $M=6\mhyphen\bar6$ in the $qq\mhyphen \bar q\bar q$ basis. For a pure $T$ state, the intrinsic Hamiltonian reads
\begin{equation}\label{eq:HN4}
 H_\text{int}=\vec{p}_x^2+\frac34\,\vec x^2+\vec{p}_y^2+\frac34\,\vec y^2+\vec p_z^2+\frac12\,\vec z^2~,
\end{equation}
and again, in a naive diquark approximation, the color factors $3/4$ are replaced by $1/2$, and this lowers artificially the energy. 

Actually, the chromo-harmonic Hamiltonian \eqref{eq:HN4} was introduced by Gavela et al. \cite{GAVELA1978459}, but in a different context. Their aim was to check the conjecture that the orbital excitations of the diquark-antidiquark system hardly decay into two mesons \cite{Chan:1978nk}, and this is, indeed, the case in this model. 
\section{Fully-heavy tetraquarks}
One can easily solve the Hamiltonian \eqref{eq:HN4}, with a ground state energy
\begin{equation}
 E_4^T=3\left(\sqrt3+1/\sqrt2\right)\simeq7.32~,
\end{equation}
which is far above the energy of two mesons, $E_\text{th}=6$ in this model. 

If one repeats the exercise with a color $M$ configuration, one finds a lower energy
\begin{equation}
 E_4^M=3\left(\sqrt5+\sqrt6\right)/2\simeq7.03~
\end{equation}
still above the energy of two mesons. If one accounts for the $T$-$M$ mixing, as done first in [..], the energy is lowered but the system remains unbound. 

Of course, the $M$ configuration cannot be understood in a naive diquark model, as the diquark is seemingly unbound, with a color coefficient $(-3/16)\langle \tilde\lambda_1.\tilde\lambda_2\rangle=-1/4$, while the exact handling of the Hamiltonian gives a potential energy $3/8(\vec x^2+\vec y^2)+\cdots$ in which the coefficient of $\vec x^2$ is positive. 

If one now solves the fully-heavy tetraquark \eqref{eq:HN} with
another potential, for instance with $v(r)=-a/r+b\,r$, $a>0$, $b>0$, one also gets the results that the  $M$ configuration is more favorable than the $T$ one, and the color-mixing does not rescue the binding. It is also found that the spin-spin force is not strong enough to produce a stable tetraquark in the fully-heavy sector.

One might wonder why the chromo-electric Hamiltonian does not bind the tetraquark, while its atomic-physics analog produces a weakly bound positronium molecule \cite{PhysRev.71.493}. An explanation has been provided in \cite{Richard:2018yrm}. Let us consider the class of models 
\begin{multline}
 h(\lambda)=\sum \vec p_i^2+(1/3+2\,\lambda)\,[v(r_{12})+v(r_{34})]\\{}+(1/3-\lambda)\,[v(r_{13})+v(r_{14})+v(r_{23})+v(r_{24})]~,
\end{multline}
where $v$ is attractive. 
After suitable numbering, it corresponds to the threshold made of two mesons or two atoms ($\lambda=1/3$), to the Ps$_2$ molecule for $v(r)=-1/r$ and $\lambda=-2/3$, to a $T$-color tetraquark for $\lambda=1/12$ and a $M$ one for $\lambda=-7/24$. In the Hamiltonian $h$, the cumulated strength of attraction is $2$, but it is spread differently among the pairs, depending on $\lambda$. For $\lambda=0$, the Hamiltonian is fully symmetric, and from the variational principle, the ground-state energy $e(\lambda)$ is higher for $\lambda=0$ than for any $\lambda\neq0$. The ground-state energy is  a concave function of $\lambda$, since $\lambda$ enters linearly the Hamiltonian~\cite{Thirring:2023497}, so  $0<\lambda_1<\lambda_2$  or $0>\lambda_1>\lambda_2$ implies $e(0)> e(\lambda_1)>e(\lambda_2)$. This demonstrates rigorously that a $T$ tetraquark cannot lie below the two meson threshold in this model. If one assumes that the energy $e(\lambda)$ is nearly symmetric, for instance parabolic, near $\lambda =0$, one gets the plausible, but not full rigorous, result that $|\lambda_1|<|\lambda_2|$ implies $e(\lambda_1)>e(\lambda_2)$, which means that Ps$_2$ is stable and the $M$-diquonium is not. Of course, the stability of Ps$_2$ is  explained differently in the textbooks on quantum chemistry, but there is no contradiction: the asymmetries in the strength parameters induces some deformation of the two positronium atoms which then can adopt a configuration which favors the attractive pairs. When calculating the tetraquark energy, one always finds the tetraquark energy close to the meson-meson threshold, so one should refrain from any unjustified approximation that could create binding artificially. In the literature, one finds sometimes too large a removal of the center-of-mass energy. Also in the cluster approximation, one replaces, say, $v(r_{13})+v(r_{14})$ by $2\,v(r_{1c})$ where $c$ is the center of the ($3,4)$ pair. If $v(r)$ is Coulombic and the $(3,4)$ system isotropic, the replacement is exact, from the well-know Gauss theorem. For a linear potential, this is an \emph{antivariational} approximation, which might lead to misleading conclusions about the stability. 

\begin{figure}[h!]
 \centering
 \raisebox{20pt}{\includegraphics[width=.15\textwidth]{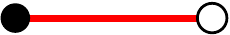}}
 \qquad
  \includegraphics[width=.12\textwidth]{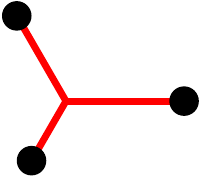}
 \caption{Schematic representation of the linear confinement for mesons (left) and baryons (right)}
 \label{fig:mes-bar}
\end{figure}

\begin{figure}[h!]
 \centering
 \raisebox{0pt}{\includegraphics[width=.16\textwidth]{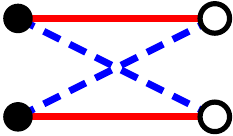}}
 \qquad
  \includegraphics[width=.16\textwidth]{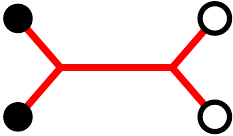}
 \caption{Schematic representation of the linear confinement for tetraquarks: flip-flop (left) and connected string (right)}
 \label{fig:tetra}
\end{figure}

The model \eqref{eq:HN} corresponds to a pairwise interaction, with a color-octet exchange. For the linear part, the $b\,r$ potential in mesons can be understood as the energy of a flux tube going straight from the quark to the antiquark. For the baryons, it has been suggested long ago that the flux tube adopt the structure of a Fermat-Torricelli $Y$-shape \cite{Artru:1974zn,Dosch:1975gf} linking each quark to a ``junction'', which is a kind of signature of the baryon number \cite{Montanet:1980te,Rossi:2016szw}, as schematically pictured in Fig.~\ref{fig:mes-bar}. 
The baryon phenomenology is not significantly changed by adopting this $Y$-shape potential instead of the $\tilde\lambda.\tilde\lambda$ rule. But for tetraquarks, new perspectives are open. The generalization of the Fermat-Torricelli string is shown in Fig.~\ref{fig:tetra}. The choice is between  the minimum of the two possible quark-antiquark links and the connected configuration with two junctions, and this provides more attraction than the $\tilde\lambda.\tilde\lambda$ ansatz, leading to some stable multiquarks, for instance $bb'\bar b\bar b'$, where $b'$ is  fictitious heavy quark, different from the $b$ quark by having the same mass or nearly the same mass. But it does not work for $bb\bar b\bar b$ with identical $b$, since the rearrangement of the flux tubes in Fig.~\ref{fig:tetra} implies changing freely the internal color wave function. 
It remains that there is some attraction between two quarkonia, possibly leading to resonances, which might explain the signal  recently seen by LHCb \cite{Aaij:2020fnh}. In the quark model, one needs dedicated techniques to describe resonances, and these techniques differ from the ones used for bound states. For instance a group has shown that for the pentaquark $\overl QQqqq$, the method of real scaling (sometimes called stabilization) enables to separate clearly the resonances from the background of states mimicking the continuum. This method, or another one, has to be applied to $cc\bar c\bar c$ configurations to confront the LHCb results. 
\section{Open-flavor tetraquarks}
Another way to fight against the instability is to introduce some favorable symmetry breaking. It is well known that any asymmetry lowers the energy of the ground state. For instance, the one-dimensional harmonic oscillator $h(\alpha)=p^2+x^2+\alpha\,x$ has an energy smaller than the energy for $\alpha=0$. This can be seen by direct calculation or by applying the variational principle, using as trial function the solution of the even part of $h(\alpha)$. This reasoning holds whenever a symmetry is broken: parity as in the above example, permutation symmetry, charge conjugation $C$, etc.\@ When one deals with few-body systems, symmetry breaking usually lowers both the ground-state energy of the whole system and  the energy of its threshold. But it often happens that the latter benefits more from the asymmetry, and thus that the stability is deteriorated by the symmetry breaking. For instance, if one breaks particle identity in $e^+e^+e^-e^-$, one observes that $M^+m^-M^+m^-$ becomes unbound if $M/m\gtrsim 2.2$ (or, of course, $M/m\lesssim 1/2.2$)~\cite{PhysRevA.57.4956,1999fbpp.conf...11V}.

The most favorable situation occurs  for the breaking of $C$, as the  threshold energy remains constant. Consider, indeed,
\begin{equation}
 H=\frac{1}{2\,M}(\vec p_1^2+\vec p_2^2)+\frac{1}{2\,m}(\vec p_3^2+\vec p_4^2)+V~.
\end{equation}
It can be decomposed as 
\begin{equation}
\begin{aligned}
 H&=H_0+H_1~,\\
 H_0&=\left(\frac{1}{4\,M}+\frac{1}{4\,m}\right)\left(\vec p_1^2+\vec p_2^2+\vec p_3^2+\vec p_4^2\right)+V~,\\
 H_1&=\left(\frac{1}{4\,M}-\frac{1}{4\,m}\right)\left(\vec p_1^2+\vec p_2^2-\vec p_3^2-\vec p_4^2\right)~.
\end{aligned} 
\end{equation}
From the variational principle, the ground state energies fulfill that $E(H)<E(H_0)$, while $H$ and its symmetric part $H_0$ have the same threshold. Thus the stability is improved if one starts form a symmetric Ps$_2$ in atomic physics, and in quark model calculations, the  stability is reached for $M/m$ large enough. The possibility of binding doubly-flavored tetraquark $QQ\bar q\bar q$ tetraquark has been suggested many years ago, confirmed by several studies, and recently revisited, though in some of the latest contributions, the early references are advertently or inavertently omitted. See, e.g., \cite{Richard:2016eis} for refs. 

For $QQ\bar u\bar d$ with $J^P=1^+$, the above chromoelectric mechanism is reinforced by a favorable interaction in the light sector. There is now a consensus that $bb\bar u\bar d$ is stable. For $cc\bar u\bar d$, a calculation by Rosina et al.\ \cite{Janc:2004qna}, later improved by Barnea et al.~\cite{Barnea:2006sd}, finds it stable, but this is somewhat model dependent. Note that in a naive diquark treatment, there is an excess of attraction due to the simplified dynamics, as already stressed, and also a lack of attraction due to the neglect of the $M$ color component, so that in some cases, there is a fortuitous cancellation of errors. 
\section{Pentaquarks}
We already mentioned the LHCb pentaquarks and the interesting attempts to describe them in a quark model with the method of real scaling.  One should stress that other flavor or spin configurations are predicted, for instance $\bar c c s qq$. In Ref.~[\cite{Richard:2017una}, it was pointed out that some states with higher spin or isospin than the LHCb pentaquarks might be bound, by a subtle cooperative effect of both chromelectric and chromomagnetic interactions. Such states should be searched for in dedicated final states that do not always involve a $J/\psi$. 
\section{Heavy hexaquarks}
Last year a very interesting lattice QCD calculation was published, claiming that the some fully-heavy hexaquarks are stable with respect to their lowest threshold for  dissociation into two baryons \cite{Junnarkar:2019equ}, for instance
\begin{equation}
 cccbbb<bbb+ccc~.
\end{equation}
This result motivated other studies. In \cite{Richard:2020zxb}, a potential model was adopted, and the six-body problem solved carefully. No bound state was found. So the question remains open, and, of course, theorists have plenty of time to debate before such very exotic states could be searched for experimentally. 

In potential model calculations, one can study how $bbbccc$ behaves when one starts from equal masses $m_b=m_c$, in which case the hexaquark is unbound,  and let $m_b$ increase and $m_c$ decrease: it is found that the hexaquark benefits less from the symmetry breaking than its threshold $bbb+ccc$. 

To repeat the reasoning that was made earlier for tetraquarks, the lack of binding for the equal-mass case $QQQQQQ$ comes from that the potential is distributed equally among all pairs, while in the threshold $QQQ+QQQ$, only 6 out of the 15 pairs experience a potential $v(r)/2$, in the notation of \eqref{eq:HN}, while the interaction is switched off for the 9 other pairs.
\section{Outlook}
In recent years, new configurations have been accessed, and the results have stimulated many interesting studies. In constituent models, the early speculations on multiquarks have been focused on the chromomagnetic part of the interaction. It is now realized that the chromoelectric part can also be a source of binding, when the masses are arranged optimally. The differences and analogies with few-charge systems in atomic physics provide a deeper understanding of the quark dynamics in exotic hadrons. 

The concept of diquark remains controversial. As an approximation to the quark model, it is definitely ruled out. As a kind of effective constituent, it provides a tempting  simplification. However, one of the appealing aspects of QCD is its non-Abelian nature, which invites to a collective dynamics where all constituents contribute. Freezing out subsystems spoils the subtlety of the binding mechanisms. 

In the course of the studies of baryons and multiquarks, and in the survey of the abundant literature, one also realizes how powerful and how forgotten is he Born-Oppenheimer approximation. For a doubly-heavy baryon $QQq$, one can derive an effective $QQ$ interaction which generates at once the lowest levels. Amazingly, the $QQ$ effective potentials of $QQq$ and $QQ\bar q\bar q$ are rather similar, though there is no pronounced antidiquark clustering of the light antiquarks.\footnote{In contrast, the $ppe^-$ and $ppe^-e^-$ Born-Oppenheimer are rather different in atomic physics, as in one of the cases, the $pp$ pair is not neutralized by the light cloud surrounding it.}\@  
This is perhaps there the true diquark effect, or quark-antidiquark symmetry. Anyhow, the almost exact identity of the $QQq$ and $QQ\bar q\bar q$ Born-Oppenheimer potentials explains the inequalities written down in 
\cite{Eichten:2017ffp}. After all, in the spectroscopy of heavy quark systems, almost everything is governed by a Born-Oppenheimer scheme. The quarkonium potential is the effective $Q\overl Q$ potential corresponding to the gluon energy at given $Q\mhyphen \overl Q$ separation, the next potential generating the hybrids~\cite{Hasenfratz:1980jv}.  When light $q\bar q$ pairs take over on the gluons, one gets a picture of the $XYZ$ mesons, or at least some of them \cite{Braaten:2014qka}. But, perhaps, open flavor exotics are even more suited for implementing Born-Oppenheimer in the dynamics.

\section*{Acknowledgments} 
%\vspace*{-0.4cm}
%
This paper is dedicated to the memory of my colleague and friend Andr\'e Martin, who taught me so much about the quark model and many other issues. I~also benefited from the longstanding collaboration of Alfredo Valcarce and Javier Vijande, and from comments on the manuscript by M.~Asghar.  I wish to thank Stephan Narison for maintaining this beautiful series of conferences in spite of the present difficulties. 
%\vspace*{-1cm}
%\vfill \eject
% \bibliographystyle{elsarticle-num}
% % \bibliography{hadronbib}
% \bibliography{qcd2020bib}
%

%
\end{document}